\documentclass[cameraready]{Interspeech}

\title{Massive Open-Vocabulary Keyword Spotting}

\author[affiliation={1}, orcid=0009-0006-5287-9717]{Leonor}{Barreiros}
\author[affiliation={1,2,3}, orcid=0009-0005-2435-1416]{Raul}{Monteiro}
\author[affiliation={1}, orcid=0000-0001-5127-9038]{Afonso}{Mendes}
\author[affiliation={1}]{Gonçalo M.}{Correia}

\address{
    $^1$ Priberam Labs, Lisboa, Portugal \\
    $^2$ Instituto Superior Técnico, Lisboa, Portugal \\
    $^3$ Instituto de Telecomunicações, Lisboa, Portugal
}

\email{\{leonor.barreiros, raul.monteiro, amm, goncalo.correia\}@priberam.pt}

\keywords{automatic speech recognition, keyword spotting, embeddings compression, contextual biasing.}

\usepackage{comment}
\usepackage{multirow}
\usepackage{hyperref}

\begin{document}

\maketitle

\begin{abstract}
    Automatic speech recognition systems have been shown to under-perform when it comes to transcribing words rarely seen in the training data, namely specialized terminology. Open-vocabulary keyword spotting, combined with contextual biasing, has been shown to mitigate this issue. However, existing systems can only handle glossaries of a few hundred terms without becoming an infeasible bottleneck. We propose a system that stores features with a memory footprint up to $128$ times smaller than a comparable baseline and allows users to process massive databases while remaining open-vocabulary. Without fine-tuning the speech recognition model, our system achieves a comparable entity recall as uncompressed solutions, even in languages not seen during training.
\end{abstract}

\section{Introduction}

Automatic speech recognition (ASR) is a task where spoken language is transcribed into text. It is crucial in several applications, from virtual assistants to live captions in various domains~\cite{health-asr, school-asr}. 
The Whisper~\cite{foundations:whisper} models, which are encoder-decoder models, are some of the most popular speech foundation models for ASR, and recent works have focused on reproducing them using public data and open-source code~\cite{owsm}.
However, most models struggle with specific terminology that was rarely seen during training~\cite{errors}. This greatly impacts the applicability of ASR in production setups where domain-specific terms are employed (e.g., healthcare and air traffic control), since correctly transcribing these keywords is essential for extracting the necessary semantic information for downstream analysis.

Contextual biasing (CB) in ASR is a technique in which generation is steered with a provided context. In this study, the context is specialized terminology in the word distribution's tail, and CB amounts to promote these otherwise missed terms. 
Many works address CB by prompting~\cite{tcpgen, cb-gpu}. The decoders of acoustic encoder-decoder models such as Whisper are audio-conditional language models, which consider previously-transcribed text during generation to improve transcription coherence. In addition to the previous context, it is possible to inject a biasing list with specific terminology directly into the ASR model decoder prompt \cite{cb-hallucinations}. To address long biasing lists, some solutions integrate keyword spotting (KWS) models to detect relevant keywords that should integrate a short biasing list \cite{learn-prompt}. Keyword spotting is a task in which a system determines whether a keyword is said in an utterance. Open-vocabulary KWS (OV-KWS) is a subclass of KWS where any keyword may be used for queries at inference time.
Given a glossary of entity words and an utterance, the ASR systems considered in this work first use an OV-KWS model to determine which keywords appear in the utterance. Then, the detected keywords are prompted into the ASR model's decoder, which will be positively biased towards predicting them.

This technique was originally proposed by~\cite{cb-whisper} for ASR systems that integrate Whisper models for speech recognition. More recently, \cite{user-cb-whisper} improved domain generalizability. Both works rely on distributed representations extracted from audios where the keywords are spoken. The latter studied the impact of using synthetic word audios, produced with either a text-to-speech (TTS) model or cut from utterances where they natually appear, whereas the former only relied on synthetic keyword audios.
Other solutions, such as \cite{latent, agreement} encode utterance audios and keyword texts and project audio and text embeddings to a shared hyperspace. 
While efficient, text-audio embeddings lose acoustic information, and the way words are pronounced is not deterministically tied to how they are written.
The importance of the information given by the acoustic encoder, to correctly transcribe rare words, was additionally documented by~\cite{errors, gen-2}.


In this work, we claim the works of \cite{cb-whisper, user-cb-whisper} have limited applicability to production environments.
The adopted acoustic model from \texttt{Whisper-medium}, which has $24$ transformer layers, is used to encode both utterance and keyword audios. In \cite{cb-whisper}, the authors chose to aggregate the representations from the $10^{th}$ to the $21^{st}$ layer and in the latter, the authors aggregated the representations from the last $12$ layers, without specifying what motivated their choices.
Most importantly, acoustic features are very high-dimensional, and, consequently, it is infeasible to process glossaries with a massive number of items using an acceptable amount of resources. 
Namely, in these works glossaries of up to $500$ keywords were used, and in our experiments we constructed a glossary of $16,062$ terms.


We propose a system that makes a significant leap in latency and memory over the baseline of \cite{user-cb-whisper} by adding an embedding compression module that projects the informative cues into a smaller-dimensional manifold. Our experiments demonstrate our system can process massive databases $6$ times faster using $128$ times less memory.
Our contributions are:
\begin{itemize}
    \item We propose an automated way to sparsely identify layers from transformer-based acoustic encoders that are most predictive for keyword spotting.
    \item We propose an embedding compression mechanism over three dimensions that is able to compress acoustic embeddings $128$ times while preserving the baseline performance.
    \item We demonstrate, in a real production system, the impact (latency- and memory-wise) of our embedding compression methodology.
\end{itemize}
The source code for this work is in \url{https://github.com/Priberam/Enhance-CB-Whisper}.

\section{Methodology}

\begin{figure}[t]
  \centering
  \includegraphics[width=\linewidth]{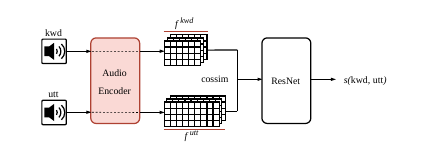}
  \caption{OV-KWS: keyword and utterance are encoded, cosine similarity matrices are computed, and a ResNet detects whether the keyword is present in the utterance.}
  \label{fig:kws-overview}
\end{figure}

\begin{figure}[t]
  \centering
  \includegraphics[width=\linewidth]{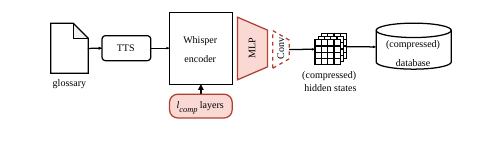}
  \caption{Embedding compression pipeline. The most relevant layers of the Whisper encoder are selected, an MLP reduces the hidden dimension, a CNN reduces the temporal resolution.}
  \label{fig:compression-diagram}
\end{figure}

\subsection{Baseline for OV-KWS}

The OV-KWS system from CB-Whisper \cite{cb-whisper} is based on two modules. \textbf{(1)} A transformer-based audio encoder from a Whisper model that encodes each audio as embeddings in $\mathbb{R}^{l \times f \times h}$, where $l$ is the number of selected transformer layers from which representations are extracted, $f$ is the number of frames, and $h$ is the hidden dimension. \textbf{(2)} A Resnet binary classifier that takes pairs of keyword and target utterance embeddings and predicts whether the keyword is or is not present in the utterance. Keyword and utterance pairs are jointly represented as $l$ cosine similarity matrices, on which the classifier attempts to detect diagonal features.  During training, positive examples are randomly sampled for each utterance and keywords not present in the utterance but lexicographically similar to the positive word are used as hard negative examples. Fixed-size cosine similarity matrices are used to improve generalization, where the utterance and keyword embeddings are zero-padded in the frame axis to $f^{utt}$ and $f^{kwd}$, respectively. The baseline OV-KWS system is schematized in Figure~\ref{fig:kws-overview}. 

In production settings, the embeddings of keywords from a glossary of words are pre-computed from audios where the keywords are spoken and then stored in a database. Keyword audios can be synthesized using a text-to-speech (TTS) model or cut from utterances where they naturally occur. This works as long as the training data includes both TTS-generated and natural-speech audios for the keywords~\cite{user-cb-whisper}. The existing solution was proven to be effective for keyword glossaries up to a few hundreds. However, since its complexity grows linearly with the number of entity words in both memory and latency, it becomes inefficient in production environments, where latency is critical, and glossaries may have thousands of keywords. 

More specifically, when storing a database of keywords in $32$-bit, using the baseline~\cite{cb-whisper}, where $l=12$, $f=150$ and $h=1024$, requires $4 \times 12 \times 150 \times 1024\approx7.3$~MB per keyword. If the number of keywords is larger than $11,650$, it is not possible to load the glossary database on a GPU with $80$GB memory (\textit{e.g.}, NVIDIA A100).
Our work explores simple techniques to improve the efficiency of KWS with respect to this baseline. While remaining proportional to the number of keywords (to preserve OV), our system's representations are $128$ times smaller, allowing $894,784$ terms on a GPU with $48$GB memory (\textit{e.g.}, NVIDIA L40). Figure~\ref{fig:compression-diagram} illustrates the process of creating a compressed keyword database.

\subsection{First compression: selecting Whisper-encoder layers}\label{sec:l-comp}

The first compression technique amounts to choosing the transformer layers from the audio encoder whose representations are aggregated to generate embeddings, both for the utterances and keywords. We propose an automated strategy to sparsely choose the layers whose representations are most predictive for KWS. Let $l$ be the number of layers of the audio encoder and $w\in\mathbb{R}^{l}$ a trainable score vector. Applying the sparsemax activation function \cite{sparsemax} yields a probability vector $p=\text{sparsemax}(w)\in\Delta^{l-1}$, where $\Delta^{l-1}$ is the $(l-1)$-dimensional probability simplex. Consider a complete embedding $x\in\mathbb{R}^{l\times f\times e}$ that aggregates the representations of all encoder layers. We gate the embedding as $\tilde{x}=p\odot x$, where $\odot$ is the Hadamard product. The score vector should be trained jointly with a Resnet classifier on these gated embeddings, using the usual binary cross-entropy loss $\mathcal{L}_{BCE}$ for KWS together with an auxiliary loss $\mathcal{L}_{aux}=H(p)$ to promote sparsity, where $H(\cdot)$ stands for the entropy of a categorical distribution. The non-zero entries of $p$ after training determine a sparse set of layers whose representations together are predictive for KWS. Denote by $l_{comp}$ the size of this set.

\subsection{Second compression: reducing the hidden dimension}\label{sec:e-comp}

Embeddings can be further compressed by reducing the hidden dimension from $h$ to $h_{comp}$, where $h_{comp}<h$. To accomplish this, a lightweight one-hidden-layer feed-forward network (FFN) can be trained to compress the embeddings along the hidden dimension, both for the utterance and keyword embeddings. The FFN and the Resnet classifier should be jointly trained with the usual binary cross-entropy loss for KWS.

\subsection{Third compression: reducing the frame rate}\label{sec:f-comp}

The embedding size depends linearly on the number of frames, $f^{utt}$ and $f^{kwd}$ for the utterances and keywords, respectively. Decreasing the frame rate is thus a natural strategy to further compress the embeddings. Whisper encoders induce a frame rate of \SI{50}{\second^{-1}} in the embeddings, which can be further reduced using a lightweight 1D convolutional neural network (CNN) along the frame direction with adequately designed kernel sizes, strides, and pooling layers. If the frame rate decreases by a factor of $\alpha$, so do the number of frames for the utterances and keywords, $f_{comp}^{utt}=f^{utt}/\alpha$ and $f_{comp}^{kwd}=f^{kwd}/\alpha$, respectively.

\subsection{Contextual biasing Whisper}

Whisper is an end-to-end ASR model composed of an encoder and a decoder. The Whisper decoder is an audio-conditional language model, trained on a next-token prediction task. Whisper was also trained to consider previous tokens during generation. This is implemented by adding context tokens before the \texttt{<|startoftranscript|>} token, which indicates the beginning of generation.
CB replaces the history tokens with a custom prompt, in this case the hotwords detected by the OV-KWS mechanism. This was originally proposed by~\cite{cb-whisper} and is equally applied in this work.

\section{Experimental setup}

\subsection{Datasets}

\subsubsection{Training data}

The training data for our models was extracted from the Multilingual Librispeech (MLS) corpus~\cite{mls}, derived from read audiobooks. 
As it is an imbalanced dataset, we reused~\cite{user-cb-whisper}'s code-base to prepare \SI{25}{\hour} of training data for six languages: English, French, German, Polish, Portuguese, and Spanish. 
To define the training keywords, the authors used spaCy~\cite{spacy} to extract the nouns and proper nouns in the transcripts and randomly selected $12,000$ terms per language. This database of keywords is fixed during training and used to extract positive and negative samples.
Our models were also trained using a mixture of natural-speech and TTS keyword audios, and the latter were generated using \texttt{edge-tts}~\cite{edge-tts}. 

\subsubsection{Validation and evaluation data}

In all validation and evaluation scenarios, we consider TTS-generated keyword audios. This is because it is usually hard to obtain natural-speech audios for domain-specific glossaries.

The MLS development split was used for validation, with $200$ keywords per language and the preparation of~\cite{user-cb-whisper}.

For evaluation, we used two open-source and one internal corpora. The former are Aishell~\cite{aishell} in Chinese and ACL6060~\cite{acl6060} in English, both with manually-annotated hotwords. Aishell was recorded in an isolated environment with a high fidelity microphone, and is in a language unseen by the KWS system. ACL6060 contains audios of live research paper presentations, which are noisier and include technical terms.
The last corpus is a set of family medicine consultations in Portuguese. 
Our system was trained on Portuguese utterances, but a set of clinical dialogues allows to evaluate the advantage of CB in a real-world application.
As we want to evaluate how our system performs on massive databases, we constructed a glossary of medical keywords from the Portuguese catalogue of clinical terms~\cite{ctc}, comprising $16,062$ terms after extracting all words and removing stopwords and numerals. 

The statistics of the evaluation corpora are in Table~\ref{tab:stats-data}.

\begin{table}[b!]
    \caption{Evaluation datasets' statistics.}
    \label{tab:stats-data}
    \centering
    \begin{tabular}{l r r r}
         \toprule
         & \textbf{Duration (min)} & \textbf{Utterances} &\textbf{Entities} \\
         \midrule
         Aishell & $76$ & $808$ & $400$\\
         ACL6060 &  $51$ & $123$ & $200$\\
         Internal & $103$ & $210$ & $16,062$\\
         \bottomrule
    \end{tabular}
\end{table}

\subsection{Experiment Configurations}

Our experiments use \texttt{Whisper-large-v2} as the acoustic encoder. Compared to the baselines of~\cite{cb-whisper, user-cb-whisper}, which use \texttt{Whisper-medium}, our backbone has $l=32$ instead of $l_{medium}=24$, and $h=1,280$ instead of $h_{medium}=1,024$. We took these differences into consideration when calculating our proposal's memory savings. We set $f^{utt}=1500$, $f^{kwd}=150$. For the classifier, we used ResNet-$50$~\cite{resnet}.

Following~\cite{user-cb-whisper}, our training used a mixture of TTS and natural-speech keywords, and lexicographic hard negatives. 
We reused the baseline's hyperparameters and a learning rate of $1\times10^{-4}$ for the added parameters.
All experiments were conducted on an NVIDIA L40 GPU.

\subsubsection{First experiment: \texttt{L}-comp}

Our first experiment only applies the layer compression introduced in Section~\ref{sec:l-comp}. We added a trainable score vector $w\in\mathbb{R}^{32}$ and trained it with a Resnet classifier on the produced gated embeddings. After training, the sparse score vector identified $3$ predictive layers: $14$, $16$ and $32$ (where $l=32$).

\subsubsection{Second experiment: \texttt{LH}-comp}

Our second experiment accumulated layer compression with the embedding compression introduced in Section~\ref{sec:e-comp}. We added an FFN with $\frac{h}{2}$ hidden units and selected $h_{comp}=64$.

\subsubsection{Third experiment: \texttt{LHF}-comp}

Our third experiment accumulated the two previous compressions with the resolution compression introduced in Section~\ref{sec:f-comp}. We added a $1$D CNN composed of a convolution operation with kernel size $3$ and stride $1$, followed by a batch normalization layer, followed by a max-pooling layer with kernel size $3$ and stride $2$. This results in a compression factor $\alpha = 2$.

\subsubsection{Contextual biasing experiments}

To evaluate the impact of our solution, we compare both transcription quality and efficiency on the evaluation corpora under three scenarios: \textbf{(1)} without biasing, \textit{No-CB}, \textbf{(2)} using the recreated setup of~\cite{user-cb-whisper}, \textit{Baseline recr.}, and \textbf{(3)} using the setup with the most compressed embeddings, \textit{\texttt{LHF}-comp}.
In setup \textbf{(2)}, as it was not possible to fit the database in memory, we batched groups of $125$ keywords for an equivalent memory footprint.

To perform CB, we prepend OV-KWS to \texttt{WhisperX}~\cite{whisperx}, based on \texttt{Whisper-large-v2}, and passed the KWS output to the \texttt{hotwords} argument of \texttt{transcribe}. 
\texttt{WhisperX} leverages voice-activity detection (VAD) to process long-form audio efficiently, which is suitable for production environments.

\begin{table*}[t!]
    \caption{ASR evaluation on Aishell-test, ACL6060-test and our internal evaluation corpus. We report mixed error rate (MER), recall (R), real time factor (RTF) and the database's memory footprint (MB).}
    \label{tab:trans-eval}
    \centering
    \begin{tabular}{l r r r r r r r r r r r r}
         \toprule
         & \multicolumn{4}{l}{\textbf{ACL6060}} & \multicolumn{4}{l}{\textbf{Aishell}} & \multicolumn{4}{l}{\textbf{Internal}} \\
         & MER & R & RTF & Memory & MER & R & RTF & Memory & MER & R & RTF & Memory \\
         \midrule
          No-CB (ours) & $27.6$ & $52.5$ & $0.03$ & $0.0$ & $18.0$ & $40.2$ & $0.03$ & $0.0$ & $29.5$ & $70.9$ & $0.07$ & $0.0$ \\
          Baseline recr.~\cite{user-cb-whisper} & $27.0$ & $54.3$ & $0.18$ & $1,406$ & $24.9$ & $59.3$ & $0.29$ & $2,812$ & $28.7$ & $72.0$ & $4.52$ & $112,929$ \\
         \texttt{LHF}-comp &  $21.9$ & $57.2$ & $0.17$ & $11$ & $14.7$ & $71.3$ & $0.10$ & $22$ & $32.4$ & $71.5$ & $0.76$ & $882$ \\
         \bottomrule
    \end{tabular}
\end{table*}

\begin{table}[t]
    \caption{KWS metrics (F1 and F1$@5$), with a $95\%$ confidence interval, on Aishell-test, ACL6060-test and the Internal corpus.}
    \label{tab:kws-eval}
    \centering
    \begin{tabular}{l r r r}
         \toprule
         & \textbf{ACL6060} & \textbf{Aishell} & \textbf{Internal}\\
         & F1 & F1 & F1$@5$ \\
         \midrule
         Baseline~\cite{cb-whisper}       & $-$    & $89.0$ & $-$\\
         Baseline~\cite{user-cb-whisper}  & $65 \pm 4$ & $71 \pm 5$ & $22 \pm 2$ \\
         \midrule
         \texttt{L}-comp                  & $37 \pm 4$ & $43 \pm 6$ & $-$ \\
         \texttt{LH}-comp                 & $63 \pm 7$ & $84 \pm 3$ & $13 \pm 1$ \\
         \texttt{LHF}-comp                & $69 \pm 4$ & $86 \pm 4$ & $28 \pm 2$ \\
         \bottomrule
    \end{tabular}
\end{table}

\subsection{Evaluation details}

\subsubsection{Keyword-spotting model evaluation}

For the open-source corpora, performance of KWS was evaluated with F1 under the optimal threshold (defined on the validation corpora). We also produce precision-recall curves. 
For our internal dataset, we relied in top-$k$ selection, with $k=5$ and report F1$@5$. $k$ is purposefully low because using CB increases the risk of hallucinations~\cite{cb-hallucinations}. 
Results are reported with a $95\%$ confidence interval, computed with bootstrapping~\cite{confidence-intervals}.

\subsubsection{Keyword-guided ASR evaluation}

We evaluate transcription with normalized mixed-error-rate (MER) and entity recall. MER is equivalent to character error rate in no-space languages and word error rate in space-based languages. Our normalization procedure corresponds to removing punctuation and lowercasing. Recall accounts recognized keywords in the transcriptions. 
The evaluated transcription is based on the one obtained with CB, after aligning it with the unbiased transcription, on the character level, using the Needleman-Wunsh algorithm~\cite{alignment}. This helps mitigate CB-inflicted hallucinations. To obtain the KWS output, we used the optimal thresholds calculated during validation for each corpus. 

We additionally measured the resource savings when using our proposed compression method. We report the real time factor (RTF) for processing each \SI{30}{\second} utterance and the memory footprint for storing each keyword database.

\section{Experimental Results}


\begin{figure}[t]
  \centering
  \includegraphics[width=\linewidth]{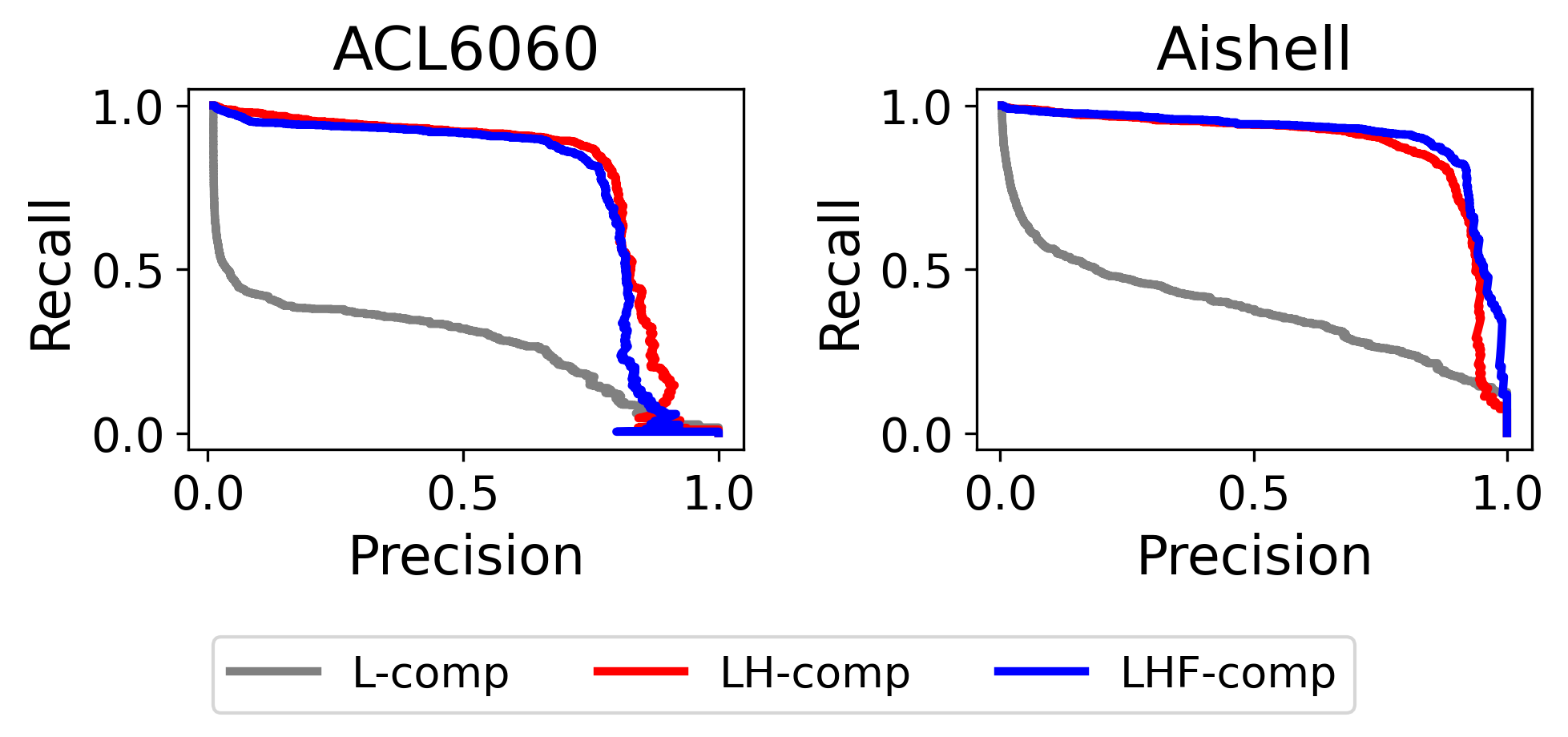}
  \caption{Precision-recall curves for the models trained with each of the proposed compressions.}
  \label{fig:pr-curves}
\end{figure}

\subsection{Keyword-spotting experiments}

The performance of our OV-KWS model and the baselines is in Table~\ref{tab:kws-eval}. 
All evaluations were out-of-domain for ours and ~\cite{user-cb-whisper}'s models, unlike for~\cite{cb-whisper} in Aishell. The model which produces the smallest embeddings, \texttt{LHF}-comp, achieves a comparable performance to the best-performing model on Aishell, even though it was not trained with this corpus or with any samples in Chinese. It also has the best F1-score on the technical ACL6060. 
Adding more trainable parameters after minimizing the number of layers allows to learn a projection of the embedding space with distinctive KWS features.

Precision-recall curves are in Figure~\ref{fig:pr-curves}. The ACL6060 test set is more challenging, we hypothesize as it contains more domain-specific terms, namely abbreviations with a non-trivial phonetic transcriptions that existing TTS systems may not capture. In addition, while Aishell was produced in a quiet environment and contains high-quality, clean audios, ACL6060 was produced with live presentations which are naturally noisier.

\subsection{Keyword-guided ASR experiments}

ASR evaluation metrics are in Table~\ref{tab:trans-eval}. On ACL6060 and Aishell, where we used threshold-based KWS on a small glossary with high recall, using CB is advantageous. 
On our Internal corpus, where KWS is performed on a massive non-curated glossary, although we see the efficiency advantage of using our method, using CB worsens ASR performance.

On the one hand, the glossaries for ACL6060 and Aishell were extracted from the nouns in their gold transcripts, so the gold keywords are aligned with the transcripts. However, for the Internal dataset, the glossary is an unprocessed set of words, including a lot of common words (\textit{e.g.}, allergens) that are distactors either because the unbiased ASR model can detect them or because they lead to false positives in CB. Whisper's prompt conditioning was trained as the previous segment's context, so it is not robust against irrelevant entries and is sensitive to spurious matches, which was also documented by~\cite{cb-hallucinations}. While our solution is the only one that allows us to efficiently process a massive glossary, additional efforts must be made to curate production databases to the desired applications.
On the other hand, analyzing the MER, \texttt{WhisperX} uses VAD for segmenting audios, which may lead to error propagation. For the open-source corpora, where utterances are short, disabling VAD would ensure no information would be lost. For our internal corpus, we had to segment audios as we used entire consultations which were much longer than \SI{30}{\second}.

Finally, we note that only with \texttt{LHF}-comp were we able to process a massive glossary. 
Whereas for the smaller, manually annotated, glossaries of ACL6060 and Aishell the RTF is acceptable for deployment, for the real-world glossary the overhead of batch processing (which implies GPU transfers) and KWS computation becomes infeasible.

\section{Conclusion}

Contextual biasing is an effective technique to improve the quality of ASR systems in specialized domains. Given a biasing list of domain-specific terms, from the tail word distribution, it steers generation to include them. 
Open-vocabulary keyword spotting aids in ensuring concise biasing lists, with terms that appear in the input query. 
This paper claims that efficient solutions, that encode keyword texts and utterance audios, are not effective, and that effective solutions, that encode keyword and utterance audios, are not efficient.
In scenarios such as the clinical, there is a massive amount of terms to identify, and KWS cannot be a bottleneck. 
We propose compressing Whisper encoder embeddings, including sparsely selecting audio encoder layers whose representations are more predictive for KWS, together with a lightweight compression module that compresses along the hidden dimension and frame length.
Our experiments demonstrate that the compressed features are as informative as the uncompressed ones, saving $128$ times memory and $6$ times latency.
Future work on this topic should explore compression on the glossary dimension, using domain-specific heuristics. It should also explore current techniques to reduce hallucinations even in the presence of distractor words.


\section{Acknowledgments}

This research was supported by the Portuguese Recovery and Resilience Plan through project C645008882-00000055 (i.e., the Center For Responsible AI).

\section{Generative AI Use Disclosure}

All references were obtained via Google Scholar search, and we studied them ourselves.
We used Claude Code (Opus 4.8) for the integration of the codebase we produced (without any generative AI) with the existing codebase for publication. No new code was generated, and manually reviewed it to ensure that.
The entirety of this paper and its revisions were written manually.

\bibliographystyle{IEEEtran}
\bibliography{mybib}

\end{document}